\documentstyle[preprint,aps]{revtex}
\tighten
\begin{document}
\draft
\title{
Hierarchical Wigner Crystal at the Edge of Quantum Hall Bar
       }

\author{ Yuli V. Nazarov}
\address{
Faculteit der Technische Natuurkunde, Technische Universiteit Delft,
 2628 CJ Delft, the Netherlands
	}
\maketitle
\begin{abstract}
 We show that quasiholes persist near the edge of
  incompressible Quantum Hall state forming a Wigner structure. The average
 density of quasiholes is fixed by electrostatics and decreases slowly
 with increasing distance from the edge.
 As we see from elementary reasoning, their specific arrangement can not be a
regular Wigner lattice
 and shows a complex hierarchical structure
 of dislocations.

\end{abstract}

It is known that a homogeneous 2D electron gas in magnetic field
may form an incompressible state if the electron density $n$
satisfies the following magic relation:
$n={p}/{q}\  {H}/{\Phi_0},$
$p,q$ being integer numbers, $\Phi_0 = 2 \pi \hbar c/e$. This
is a physical origin of Integer and Fractional Quantum Hall
effects.\cite{QuantumHall}

In reality, 2D gas structures are finite. Their size is usually controlled
by means of metal gate electrodes imposed on the top of the heterostructure
and biased at large negative potential. The electrodes deplete the
2D underneath and thus shape it. 2D gas is not homogeneous near the edge.
The structure of the edge of the incompressible state
always receives much attention.\cite{edge}
The origin of this is
that the electric
transport is believed to take place only near the edge.
Early studies neglected interelectron electrostatic interactions.
In this approach, the edge of incompressible state is very sharp being smoothed
at
distances of the order of magnetic length $l_H \equiv \sqrt{\hbar c/eH}$.
Owing to the magic relation, $\l_H$ is of the order of interelectron distance.
Lately it has been recognized that the macroscopic electrostatics
is of a crucial importance for the problem. Powerful electric
forces dilute the 2D near the edge at distances much larger than $\l_H$.
Moreover, a small deficit of the electron density persists far from the edge.
This concept was first introduced in \cite{McEuen} and has been given a refined
formulation in \cite{Shklovskii}.

In present study we concentrate on the region where the deficit of the
electron density is small compared to $n$.
We argue that the deficit is due to discrete quasiholes.
We regard those as classical particles with Coulomb interaction.
The arrangement of the quasiholes must be close to Wigner crystal.
However, it is not a regular lattice.
 We argue that the actual arrangement is a hierarchical
crystal of dislocations in Wigner lattice. The arguments given
are very straightforward and appear to be of a general nature.

Let us first recall the electrostatic picture of the Hall bar
edge.\cite{Shklovskii}
We restrict ourselves to a simplest geometry (Fig.1). 2D gas  and the  gate
electrode
are assumed to be in the same plane $z=0$, the density and voltage distribution
are uniform in $y$ direction.
The gate electrode is at $x<0$, the 2D gas density reaches is
unperturbed value $n(\infty)$ at $x \rightarrow \infty$.
Let $a$ be a separation between 2D gas and the gate.
The density profile is given by
$n(x)=n(\infty) \sqrt{\frac{x-a}{x}}.$
The separation is determined by the voltage difference $V_g$ between
the gate and the 2D gas, $a = V_g \epsilon (2 \pi^2 n(\infty)e)^{-1})$.
$\epsilon >>1$ being the dielectric constant of the host material.
As it has been argued
in \cite{Shklovskii}, this density profile remains practically
unchanged if one takes into account the inner properties of 2D gas.
The reason  is that the inner energies characterizing
the 2D gas such as Fermi energy, Landau level separation,  characteristic
Coulomb energy $e^2 \sqrt{n}/\epsilon$ are much smaller than the gate
voltage $V_g$.
Thus nothing in 2D can withstand the external electrostatic forces
and provide  sensible deviations from the profile.

Let us consider an asymptotic region $x >> a$. In this region the density
deficit has a long power-law tail
\begin{equation}
n(\infty)-n(x) = \frac{a}{2x} \equiv \frac{V_g \epsilon}{4 \pi^2 e x}
\label{tail}
\end{equation}
We can verify this peculiar shape of the tail considering
the electrostatic problem simplified to a primary school level.
Let us consider  it at the scales larger than $a$. The potential
distribution can be readily found from boundary conditions
$V(x)=0 \  {\rm at}\  x>0, \ V(x)=V_g \  {\rm at}\  x<0 \  {\rm if}\  z=0.
$
The equipotentials  are just rays beginning in the origin,
and $V(z,x)= 2V_g/\pi \vert \arctan(x/z) \vert$. Derivative of $V(x,z)$
with respect to $z$ at $z=0, x>0$ gives an electric field perpendicular
to 2D plane and, consequently, the density deficit induced by the gate.
This, of course, coincides with (\ref{tail}).

If $n(\infty)$ corresponds  to a metallic, compressible state
there is no difficulty to change its density and  provide a small density
deficit.
Very different situation takes place if $n(\infty)$ corresponds to
an incompressible state satisfying the magic relation.
By virtue of incompressibility, the density deficit must be dumped
into positively charged excitations on the background of the incompressible
state(quasiholes). The nature of these excitations depends on the type
of incompressible state: for IQHE, those are just common holes with elementary
charge.
For FQHE incompressible state, the quasiholes bear a fractional charge
$e/q$. Electrostatic interaction between holes can not be screened by the
incompressible background, so they repel each other as common charges do.
As common charges, the quasiholes are affected by magnetic field
that localizes their wave functions at space scale of the order of $l_H$.
If the density deficit is small enough, the average distance between
quasiholes is much larger than magnetic length. Due to repulsion,
they never come close to each other, their wave function do not
overlap and quantum mechanical effects are negligible. Therefore
at low densities the quasiholes can be viewed
as classical particles subjected to Coulomb repulsion, exactly like
electrons forming the common Wigner structure.\cite{Fertig2}
Very similar situation in quantum dots has been studied in \cite{Molecule}.

	At larger densities quasihole wave functions are forced to overlap
and they form a collective metallic state.
So far no theory gives  a quantitative description of this transition.
To estimate the critical concentration, one can draw an analogy with
the low density electron gas where such transition occurs at
$n_c \approx 0.2 H/\Phi_0$.\cite{QuantumHall} Summarizing all that, we shall
accept the
following picture  of the  Quantum Hall bar edge (Fig. 1): the
area adjacent to the edge can be schematically separated into two regions.
 In the region I, the density deficit is big enough to provide the metallic
state. This region may be subdivided into several disconnected metallic
regions by dipolar strips corresponding to incompressible states with
density lower than $n$. \cite{Shklovskii} In the region II, we have
the incompressible state doped by distinct quasiholes. The density of
dopants drops according to Eq. \ref{tail} as
$n_0 \equiv q (n(\infty)-n(x))=\frac{V_g q \epsilon}{4 \pi^2 e x}
\equiv \frac{1}{b x}$.
A good estimation for $a,b$ would be $b = 3 {\rm nm}$, $a=
400 {\rm nm}$.

Now we are in position to analyze what is the equilibrium arrangement
of the quasiholes far in the region II, at $x>>a$.
In the following consideration we shall completely neglect
the potential disorder. It may be very important in the systems of the kind
since the quasiparticles can be pinned in random potential
minima. In a realistic structure, random potential landscape is created
by donors spatially separated from 2D gas. The Wigner structure
survives provided the separation exceeds the average distance
between (quasi)particles. Since there are no fundamental reasons
limiting the separation the structures may be in principle as
clean as required.
Anyway, it seems to be logical to comprehend first
the situation in the absence of disorder.

Let us recall that we now deal with classical particles. They repel
each other and their average density is fixed by external means.
If the fixed average density were uniform the particles would form
a triangular Wigner lattice. We shall expect that their {\it local}
arrangement is close to the one if the density varies slowly.
However, it is easy to see that the {\it global} arrangement
can not be a regular lattice of any kind. The only way to change
the density in a regular lattice is to stretch it in $x$ direction.
This procedure does not preserve the local triangular arrangement of
the particles and the resulting stricture would have been extremely
energy unfavorable. In terms of elasticity theory,\cite{elasticity}
there would be an enormous stress in the lattice.
There should be some defects in the lattice to relieve
the stress and to restore the favorable local arrangement.

The minimal energy configuration is provided by point-like defects:
dislocations.
The triangular local arrangement will be distorted for the particles
only within several lattice spacings from a dislocation and will be preserved
for the most of them. The density of dislocations can be evaluated
from the following consideration. Let us suppose that the local
triangular arrangement is realized for most of the particles.
The number of rows (extended in $x$ direction) per unit $y$ length
is then given by
$N_{row} = \beta \sqrt{n_0}$,
$\beta$ being lattice-dependent dimensionless coefficient,
$\beta= 2/3^{3/4}$ for triangular lattice.
Each dislocation is known to swallow one row \cite{elasticity}, so that
the dislocation density is just a derivative of the number of
rows with respect to $x$:
\begin{equation}
n_1= -\frac{d}{dx}N_{row}= -\beta \frac{d}{dx} \sqrt{n_0} =
\frac{\beta}{2 \sqrt{b x^3}}
\label{dislocationdensity}
\end{equation}
We see that $n_1 << n_0$ in the region of interest, this validates
the scheme we use.

We seem to advance in understanding of particle arrangement. The question
remains what is
the {\it dislocation} arrangement. They also can be viewed as particles,
and their average density is fixed by the relation (\ref{dislocationdensity}).
As we know from elasticity theory,\cite{elasticity} they
repel each other.
 So we come back to the initial problem
with changed fixed density! We have to repeat the procedure
considering dislocations in the lattice of dislocations, dislocations
in the lattice of dislocations in the lattice of dislocations and so on.

We see that a sophisticated hierarchical structure is realized
in the system under consideration. There is an infinite
number of "particle" generations, the "particles" of $m$th
generation being dislocations in the lattice of the particles
of $m-1$ generation. The "particle" density for each generation
is given by \cite{footnote}
\begin{equation}
n_m= - \beta \frac{d}{dx} \sqrt{n_{m-1}} = \frac{A_m}{\beta^2 x^2}
(\frac{\beta^2 b}{x})^{2-2^{-m}},
\end{equation}
\[ A_m = \prod^{m}_{j=1} (1-2^{-j})^{2^{j-m}}\]
The densities of high generation particles quickly approach the
limiting function $n_\infty= \beta^2/x^2$. For the limiting distribution,
the notion of lattice no longer makes sense
since the average interparticle distance is of the order of $x$.
This also implies that the difference between generations becomes
less pronounced with growing $m$. Owing to this, only a finite
number of generations can be actually observed in a realistic
system. This depends on the system size: the larger the size,
the more generations can be resolved.
The  estimations for a Quantum Hall bar show that
at realistic submillimeter scale one can see a lattice formed
by forth generation "particles" and possibly distinct "particles" of fifth
generation.

A small segment of the resulting arrangement is presented in Fig.2.
It is an illustration, no attempt to minimize the structure energy has
been performed. For simplicity I assumed that the dislocations
shift the particles in $y$ direction only. The magnitude of the shift
is proportional to the bearing at which the particle is seen from the
dislocation location. Three generations are
visible in Fig. 2: quasiholes, dislocations in quasihole lattice
and a dislocation in the dislocation lattice.

What may be the experimental manifestations of the hierarchical crystal?
Since the quasiparticles forming the structure are localized they
do not contribute to electric current and hardly can be seen in dc
transport measurements that are so common in Quantum Hall devices.
It seem  feasible \cite{Pothier}
to measure directly the charge distribution over the bar by highly
sensitive Coulomb blockade electrometer. Another possibility is to study
microwave response of the heterostructure: the oscillatory modes
of the hierarchical crystal can be revealed in this way, similar to
those ones of a common Wigner crystal. \cite{Williams}

The arguments given are so evident and general that the scheme proposed shall
not
be supported by any analytical or numerical calculations. Besides,
such calculations are very difficult to perform
so it remains a challenging task for the future.

There is an intrinsic difficulty with numericals arising
from the simple fact that
 any arrangement of dislocations corresponds to
a local energy minimum. Thus they are difficult to anneal, and the absolute
minimum is difficult to find. The state of the art of the Wigner numerics
one can find in \cite{Fertig1,Fertig2}. Due to the difficulties described,
the number of particles considered is at most of the order of hundred.
It makes it very difficult to get rid of edge effects.
Nevertheless, the very beginning of the hierarchical scenario can bee seen
from these numerical results. Figures 4a, 4b of \cite{Fertig2}
nicely demonstrate how the dislocations appear in the structure.

I believe that such calculations are worth doing with much greater
number of particles, despite enormous CPU time required.
The reason is a general nature of the hierarchical ordering
described. The above reasoning can be applied to any system of classical
particles with repulsion provided the external forces fix their average
density.
There are various physical realizations of this
such as Abrikosov vortices in superconducting film subjected
to external magnetic field, charged dopants in semiconductors
placed in external nonuniform electric field, electrons on
liquid helium surface in nonuniform field and others.
So I conclude stressing  that the hierarchical ordering
may occur in many instances.

I am indebted to L. S. Levitov, L.S. Glazman
and many others
for very instructive discussions of the results.
This work is a part of the research program of the "Stichting voor
Fundamenteel Onderzoek der Materie"~(FOM), which is  financially
supported from the "Nederlandse Organisatie voor Wetenschappelijk Onderzoek"
{}~(NWO).

\begin{figure}
\caption{
The structure of a Quantum Hall bar edge. The density profile is presented.
2D gas can be schematically subdivided into two regions: metallic state
(i), incompressible state doped by quasiholes (ii).
}
\label{fig1}
\end{figure}

\begin{figure}
\caption{
The segment of a hierarchical crystal. Dots correspond to quasiholes.
Dislocations in quasihole lattice are marked by diamonds. The
cross marks the dislocation in the dislocation lattice:
the second generation "particle".
}
\label{fig2}
\end{figure}

\begin{references}
\bibitem{footnote} Here I assumed that all generations form a triangular
lattice. If this is not the case, $\beta$'s will be different for different
for different generations albeit the power law exponents remain the same.
\end{references}
\end{document}